\documentclass{article}
\usepackage[margin=1in]{geometry}
\usepackage{amsmath}
\usepackage{setspace}
\usepackage{graphicx}
\usepackage{epstopdf}

\title{Covariate adjustment and prediction of mean response in randomised trials}

\newcommand{\etal}{\textit{et al }}
\newcommand{\Var}{\mbox{Var}}
\newcommand{\Cov}{\mbox{Cov}}
\newcommand\ci{\perp\!\!\!\perp}

\author{Jonathan Bartlett \\ jwb133@googlemail.com \\ Statistical Innovation Group \\ Advanced Analytics Centre \\ AstraZeneca, Cambridge, UK}

\begin{document}
\maketitle

\large

\abstract{Analyses of randomised trials are often based on regression models which adjust for baseline covariates, in addition to randomised group. Based on such models, one can obtain estimates of the marginal mean outcome for the population under assignment to each treatment, by averaging the model based predictions across the empirical distribution of the baseline covariates in the trial. We identify under what conditions such estimates are consistent, and in particular show that for canonical generalised linear models, the resulting estimates are always consistent. We show that a recently proposed variance estimator underestimates the true variance when the baseline covariates are not fixed in repeated sampling, and provide a simple adjustment to remedy this. We also describe an alternative semiparametric estimator which is consistent even when the outcome regression model used is misspecified. The different estimators are compared through simulations and application to a recently conducted trial in asthma.}

\section{Introduction}
Analyses of randomised clinical trials often adjust for one or more baseline covariates, typically through fitting a regression model for the outcome variable, conditional on randomised group and the specified baseline covariates. The purposes of such adjustment is usually primarily to take account of chance imbalance in the distribution of the baseline covariates between treatment groups. Statistically such adjustment generally improves the power of a test of the null hypothesis of no treatment effect. For linear outcome models the estimated treatment effect is also more precise, while for non-linear outcome models the situation is more nuanced, since in general covariate adjustment alters the conditional treatment effect being targeted \cite{Tsiatis2008}.

In addition to reporting baseline covariate adjusted estimates of treatment effect, trial reports almost always report crude or raw summary statistics such as means for the outcome, by treatment group. Since the primary treatment effect estimate is adjusted for baseline covariates, a natural although less common next step is to report estimates or predictions of mean outcome by treatment group which take the baseline covariates into account.

As described recently by Qu and Luo \cite{Qu2015}, some statistical packages provide estimates by calculating predictions for the mean outcome for a given treatment, setting the baseline covariates to their mean values in the sample `prediction at the mean'. An alternative, also described by \cite{Qu2015}, involves predicting the mean outcome assuming assignment to a particular treatment for each patient in the trial, using each patient's observed baseline covariates, and averaging these predictions across all patients in the trial. We refer to this as the `marginal mean' method. For linear models, the two approaches give identical results, while for non-linear models, such as logistic or Poisson regression, in general they do not \cite{Lane1982}. Indeed, for non-linear models, the population parameters being estimated by the two approaches differ.

The marginal mean approach predicts what the mean outcome would be if the population were assigned a given treatment. This is the same population parameter estimated by the crude treatment group specific mean outcome. In contrast, `prediction at the mean' predicts the mean outcome for a patient with baseline covariate values equal to their means. As noted by others \cite{Muller2014}, when some of the covariates are categorical, such as gender or region, `prediction at the mean', or prediction with other proportions assigned to the levels of categorical variables, arguably makes little sense. This is because any given patient can only belong to one level of each categorical covariate, and the resulting predicted value applies to a patient with covariate values that are never and can never be seen in the population.

Qu and Luo recently proposed using the marginal mean approach in randomised trials to estimate treatment group specific means, and advocated its use in preference over `prediction at the mean' for non-linear outcome models \cite{Qu2015}. In addition to point estimation, Qu and Luo described a variance estimator for the marginal mean estimates, and demonstrated that the resulting estimator was unbiased and confidence intervals achieved their nominal coverage level.

In this paper we investigate marginal mean estimation in further detail. In Section \ref{muhat1sec} we review crude marginal mean estimation in trials, whereby baseline covariates are not utilised. In Section \ref{baselineadj} we consider the estimator described by Qu and Luo in detail. We first describe conditions under which it is consistent, and show that for certain outcome model types it remains consistent even when the outcome regression model is misspecified. We then show that when, as is typically the case, baseline covariates would not be fixed in repeated samples, the variance estimator described by Qu and Luo underestimates the true sampling variance, and we give a simple adjustment which remedies this. In Section \ref{robust} we consider an alternative semiparamtric baseline adjusted estimator which is guaranteed to be unbiased irrespective of whether the outcome model is correctly specified. Next in Section \ref{alternativeRand} we consider the impact of stratified randomisation on the preceding results. In Section \ref{simulations} we report simulation results comparing the different estimators. As an illustrative example in Section \ref{illustrative} we present results from a recently conducted trial in asthma. We conclude with a discussion in Section \ref{discussion}.

\section{Marginal mean estimation ignoring covariates}
\label{muhat1sec}
Consider a randomised trial where $Z$ is a variable recording the an individual patient's randomised treatment group. Let $Y$ denote the patient's outcome on their assigned treatment. Let $Y^z$ denote the potential outcome for an individual patient were they, possibly contrary to fact, given treatment $z$ \cite{Hernan2017}. The (true) marginal mean outcome under treatment $z$ is then defined as
\begin{eqnarray}
\mu(z) = E(Y^z)
\end{eqnarray}
Randomisation ensures that $\mu(z)=E(Y^z)=E(Y|Z=z)$, so that $\mu(z)$ can be estimated by the sample mean outcome in those patients randomised to $Z=z$:
\begin{eqnarray*}
\hat{\mu}_{1}(z) = \frac{\sum^{n}_{i=1} 1(Z_{i}=z) Y_{i}}{\sum^{n}_{i=1} 1(Z_{i}=z)}
\end{eqnarray*}
where $1()$ denotes the indicator function which takes value one when its argument is true, and takes value zero otherwise. The mean for each treatment group is calculated and reported as a matter of course in randomised trials. Provided the trial used simple randomisation, the variance of $\hat{\mu}_{1}(z)$ can be estimated in the usual way for the variance of a sample mean of independent random variables.

In trials where the outcome $Y$ represents the number of events that occur in each patient during follow-up, the length of follow-up sometimes varies between patients (for various reasons). Let $T$ denote an individual patient's follow-up time. In this case, one typically assumes that $E(Y|T,Z=z)=\mu(z) T$, and interest lies in estimation of the event rate $\mu(z)$. In this case $\mu(z)$ is estimated by the sample rate
\begin{eqnarray}
\hat{\mu}_{1}(z) = \frac{\sum^{n}_{i=1} 1(Z_{i}=z) Y_{i}}{\sum^{n}_{i=1} 1(Z_{i}=z) T_{i}}
\label{cruderate}
\end{eqnarray}
A sandwich variance estimator for the latter estimator, derived later in Section \ref{robust}, is given by
\begin{eqnarray}
\hat{\tau}^{-2}_{z} n_{z}^{-2} \sum^{n}_{i=1} \left[1(Z_{i}=z)\{Y_{i}-\hat{\mu}_{1}(z)\}  \right]^{2} 
\label{ratemuhat1var}
\end{eqnarray}
where $n_{z}$ denotes the number of patients randomised to $Z=z$ and $\hat{\tau}_{z}$ denotes the mean follow-up time in those randomised to $Z=z$.

\section{Baseline adjusted marginal mean estimation}
\label{baselineadj}
As described in the introduction, trials almost always collect baseline covariates $X$ which are related, to a greater or lesser extent, to the outcome $Y$. In this section we consider estimation of $\mu(z)$ where we utilise such covariates in the estimation of $\mu(z)$.

\subsection{Outcome model specification and estimation}
We assume that the trial's analysis plan specifies a regression model for $Y$ with $X$ and $Z$ as covariates. This could be a fully parametric model, or a semiparametric model, but we require that it at least specifies how the mean of $Y$ depends on $X$ and $Z$:
\begin{eqnarray}
E(Y|X,Z) = h(X,Z,\beta)
\label{omCondMean}
\end{eqnarray}
for an assumed function $h(X,Z,\beta)$ and where $\beta$ are the model parameters. The most common specification is that
\begin{eqnarray*}
h(X,Z,\beta) = g^{-1}(\beta_{0}+\beta^{T}_{X} X + \beta_{Z} Z)
\end{eqnarray*}
for some link function $g()$. In the case of more than two treatment groups, $Z$ would be replaced in the linear predictor by a vector of indicator functions. The outcome model may sometimes contain additional parameters $\eta$, such as the residual variance in the case of linear regression or the dispersion parameter in the case of a negative binomial regression.  In the case where patients are followed up for different lengths of time, one assumes $E(Y|X,Z,T)=T h(X,Z,\beta)$. In statistical software this assumption is specified by passing the patient's follow-up time as an appropriate offset.

We let $\hat{\beta}$ denote the estimated value of $\beta$. In cases where the outcome model is fully parametric, the parameters are usually estimated by maximum likelihood. More generally, we assume that $\beta$ is estimated as the value solving a set of estimating equations, which are consistent for the true value of $\beta$ when the outcome model is correctly specified. We let $\tilde{\beta}$ denote the large sample probability limit of the estimator $\hat{\beta}$. This is the value that is consistently estimated as the sample size $n$ tends to infinity.

\subsection{Estimation of $\mu(z)$}
To estimate $\mu(z)$ using the baseline adjusted outcome model, we can use the law of total expectation to express
\begin{eqnarray}
\mu(z) = E(Y^z) = E\{E(Y^z|X)\}
\end{eqnarray}
This expression says that $\mu(z)$ can be calculated by predicting the mean outcome under treatment $z$ for each patient given their baseline covariates $X$, and then averaging these predictions across all patients. This motivates the estimator described by Qu and Luo \cite{Qu2015}, which is given by:
\begin{eqnarray}
\hat{\mu}_{2}(z) = n^{-1} \sum^{n}_{i=1} h(X_{i},z,\hat{\beta}) \label{muhat2}
\end{eqnarray}
We emphasise that this estimator, unlike $\hat{\mu}_{1}(z)$, uses data from all patients, and not only those randomised to $Z=z$. It is an example of a standardisation estimator, where we are standardising to the empirical distribution of the baseline covariates of all patients in the trial \cite{Hernan2017}.

\subsection{Consistency}
Since $\hat{\mu}_{2}(z)$ is the sample mean of $h(X,z,\hat{\beta})$, it follows that $\hat{\mu}_{2}(z)$ is a consistent estimator of $E\{h(X,z,\tilde{\beta})\}$. It will therefore be consistent if and only if $E\{h(X,z,\tilde{\beta})\}=\mu(z)$. A sufficient condition for this to hold is that $E(Y|X,Z)=h(X,Z,\tilde{\beta})$. This is satisfied in particular if the outcome model is correctly specified. In general, if the outcome model is misspecified, $\hat{\mu}_{2}(z)$ is biased for $\mu(z)$.

There are however important cases where $E(Y|X,Z)=h(X,Z,\tilde{\beta})$ even when the outcome model is misspecified in certain respects. Specifically, when the outcome model is a generalised linear model, it is sufficient that the conditional mean $E(Y|X,Z)=h(X,Z,\beta)$ is correctly specified. For example, if the outcome model is Poisson regression, provided the conditional mean specification is correct, there is no need for $Y$ to be Poisson distributed conditional on $X$ and $Z$. A similar argument applies in the case of negative binomial regression (NB2), whereby the regression parameters are estimated (asymptotically) without bias provided the conditional mean function is correctly specified \cite{Cameron2013}.

Moreover, in the special case of canonical generalised linear models (GLM), we show in Appendix \ref{muhat2bias} that $\hat{\mu}_{2}(z)$ remains consistent even if the conditional mean model is misspecified. This robustness property for canonical GLMs in randomised trials was previously derived within the framework of targeted maximum likelihood by Rosenblum and van der Laan \cite{Rosenblum2010}.

Consider again the setting where the outcome $Y$ represents the number of events that occur for each patient during follow-up, with $T$ denoting a patient's follow-up time. Suppose that Poission regression is used for the outcome model, with $\log(T)$ as an offset. In Appendix \ref{muhat2bias}, we show that provided $T \ci X | Z$, $\hat{\mu}_{2}(z)$ is consistent for $\mu(z)$ irrespective of whether the Poisson regression is correctly specified.

\subsection{Variance}
We now consider the variance of $\hat{\mu}_{2}(z)$, and estimators of this variance. Qu and Luo \cite{Qu2015} proposed a variance estimator for $\hat{\mu}_{2}(z)$ using the delta method, in which they conditioned on the observed covariate values $\underline{X}=\{X_{i}, i=1,..,n\}$, $\underline{Z}=\{Z_{i}, i=1,..,n\}$. The marginal mean estimator given by equation \eqref{muhat2} is a non-linear transformation of $\hat{\beta}$, and so by the delta method its (conditional) variance can be estimated by
\begin{eqnarray}
\widehat{\Var}(\hat{\mu}_{2}(z)|\underline{X},\underline{Z}) = \hat{G}_{\beta} \widehat{\Var}(\hat{\beta}|\underline{X},\underline{Z})  \hat{G}_{\beta}^{T} \label{QuLuoVar}
\end{eqnarray}
where
\begin{eqnarray}
\hat{G}_{\beta} = n^{-1} \sum^{n}_{i=1} \frac{\partial h(X_{i},z,\hat{\beta})}{\partial \beta^{T}}\label{hatgbeta}
\end{eqnarray}
For $\widehat{\Var}(\hat{\beta}|\underline{X},\underline{Z})$, one could use the model based variance covariance matrix, or as Qu and Luo also described, a robust sandwich variance estimator \cite{Qu2015}.

Qu and Luo's variance estimator for $\hat{\mu}(z)$ treats the covariate values in the sample as fixed known constants. This is in agreement with conventional inference methods for regression models parameters, which also treat the observed covariate values as fixed. In the vast majority of late phase randomised trials, the individual covariate values are not fixed, in the sense that if the sample or trial were to be repeated, the covariate values which would be observed would change.

Although not often made explicit, for inference regarding parameters of regression models, the practice of treating covariates as treated as fixed even when in truth they are not can be justified through the concept of ancilliarity. Specifically, in the regression setting, the marginal distribution of the covariates is ancilliary, which means inference for parameters of the conditional distribution of the outcome given the covariates can be performed as if the covariates were in fact fixed \cite{Buja2014}.

This argument does not apply however to $\hat{\mu}_{2}(z)$, since it depends not only on the regression coefficients of the outcome model but also on the marginal distribution of the baseline covariates $X$. In Appendix \ref{appVariance}, we use estimating equation theory to derive an expression for the repeated sampling variance of $\hat{\mu}_{2}(z)$ which accounts for $X$ being randomly sampled, and which shows that the variance estimator described by Qu and Luo is in general invalid when $X$ are not fixed in repeated sampling. In Appendix \ref{appVariance} we show that when the outcome model is correctly specified, or is a GLM or negative binomial regression with the conditional mean function correctly specified, the Qu and Luo variance estimator is downwardly biased by
\begin{eqnarray*}
n^{-1} \Var \left[ h(X,z,\tilde{\beta})- \mu(z) \right]
\end{eqnarray*}
The downward bias can thus be remedied by adding
\begin{eqnarray}
n^{-2} \sum^{n}_{i=1}  \{ h(X,z,\hat{\beta}) - \hat{\mu}_{2}(z) \}^{2} \label{quluo_adjust}
\end{eqnarray}
to the Qu and Luo variance estimator given in equation \eqref{QuLuoVar}. The form of this expression shows that the Qu and Luo variance estimator may be expected to be downwardly biased to a greater extent when the baseline covariates $X$ are strongly associated with $Y$. Conversely, we expect the Qu and Luo variance estimator to be approximately unbiased for the variance of $\hat{\mu}_{2}(z)$ when the baseline covariates are only weakly associatied with outcome.

The covarate adjusted estimator $\hat{\mu}_{2}(z)$ takes into account the effects on the outcome mean of any chance imbalances in the distribution of the baseline covariates $X$ between treatment groups. One might therefore expect $\hat{\mu}_{2}(z)$ to give a more precise estimate of $\mu(z)$ than the unadjusted estimator $\hat{\mu}_{1}(z)$. In Appendix \ref{muhat2eff} we give a sketch proof that shows that when the outcome model is a correctly specified parametric model, $\hat{\mu}_{2}(z)$ is indeed more efficient than $\hat{\mu}_{1}(z)$.

\section{Robust baseline adjusted marginal mean estimation}
\label{robust}
As described previously, the estimator $\hat{\mu}_{2}(z)$ can give a more precise estimate of $\mu(z)$ than $\hat{\mu}_{1}(z)$, by exploiting the baseline covariates and the randomisation assumption. However, it is only consistent for $\mu(z)$ in general when the outcome model used is correctly specified. In contrast, the crude mean estimator $\hat{\mu}_{1}(z)$ which does not use the baseline covariates, is (assuming there is no missing data) always unbiased.

As we describe in Appendix \ref{appmuhat3}, using the semiparametric theory developed by Tsiatis \cite{Tsiatis:2006} and Zhang \etal \cite{Zhang2008}, we can construct an estimator $\hat{\mu}_{3}(z)$ that exploits $X$ to improve efficiency, yet like $\hat{\mu}_{1}(z)$ is guaranteed to be consistent. This estimator is given by
\begin{eqnarray}
\hat{\mu}_{3}(z) &=& \hat{\mu}_{1}(z) - n^{-1} \sum^{n}_{i=1} \left[ \frac{1(Z_{i}=z)-\hat{\pi}_{z}}{\hat{\pi}_{z}} h(X_{i},z,\hat{\beta}) \right]
\label{muhat3}
\end{eqnarray}
where $\hat{\pi}_{z}$ denotes the sample proportion of patients randomised to $Z=z$ and $h(X,Z,\hat{\beta})$ denotes a working model prediction for $E(Y|X,Z)$. This estimator remains consistent even if the working model $E(Y|X,Z)=h(X,Z,\beta)$ is misspecified, since the second term in the expression has mean zero due to independence of $X$ and $Z$. However misspecification will affect the estimator's variance, with the variance being minimized when the working model is correctly specified. In Appendix \ref{appmuhat3} we show that the variance of $\hat{\mu}_{3}(z)$ can be estimated by
\begin{eqnarray}
\hat{\pi}^{-2}_{z} n^{-2} \sum^{n}_{i=1} \left[1(Z_{i}=z) \{Y_{i}-\hat{\mu}_{3}(z)\} - \{1(Z_{i}=z)-\hat{\pi}_{z}\}\{h(X_{i},z,\hat{\beta})-\hat{\mu}_{2}(z)\}\right]^{2}  \label{muhat3estvar}
\end{eqnarray}
This approach can be used for estimation of $\mu(z)$ for each treatment group. A separate model for $E(Y|X,Z=z)$ for each $z$ could be fitted to those randomised to $Z=z$, or alternatively, a single regression model could be fitted to all treatment groups, and the model fit used to construct estimates of $E(Y|X,Z=z)$ for each value of $z$. Note that if one chooses $h(X,Z,\beta)=0$, $\hat{\mu}_{3}(z)$ reduces to $\hat{\mu}_{1}(z)$, and as such, equation \eqref{muhat3estvar} can also be used to obtain a robust sandwich variance estimator for $\hat{\mu}_{1}(z)$ by ignoring the second term in square brackets (since if $h(X,Z,\beta)=0$, $\hat{\mu}_{2}(z)=0$).

In the case where $Y$ denotes the number of events observed on a patient and $T$ denotes their follow-up time, in Appendix \ref{appmuhat3} show that $\hat{\mu}_{1}(z)$ remains as per equation \eqref{muhat3}, with $h(X,z,\beta)$ denoting the working model prediction for $E(Y|X,Z,T=1)$. We also show that its variance can be estimated by
\begin{eqnarray}
\hat{\pi}^{-2}_{z} \hat{\tau}^{-2}_{z} n^{-2} \sum^{n}_{i=1} \left[1(Z_{i}=z)\{Y_{i}-\hat{\mu}_{3}(z)\} - \hat{\tau}_{z} \{1(Z_{i}=z)-\hat{\pi}_{z}\}\{h(X_{i},z,\hat{\beta})-\hat{\mu}_{2}(z)\} \right]^{2} 
\label{muhat3estvarfup}
\end{eqnarray}
where $\hat{\tau}_{z}$ denotes the mean follow-up time in those randomised to $Z=z$.

In Section \ref{baselineadj} we described how $\hat{\mu}_{2}(z)$ is consistent irrespective of whether the outcome model is correctly specified in the special case that the latter is a canonical generalised linear model. This result can also be demonstrated in this special case by showing that $\hat{\mu}_{2}(z)=\hat{\mu}_{3}(z)$, and hence $\hat{\mu}_{2}(z)$ inherits the robustness property of $\hat{\mu}_{3}(z)$.

\section{Stratified randomisation}
\label{alternativeRand}
Thus far we have assumed that each patient's treatment group is assigned using simple randomisation, such that the data on each patient is independent and identically distributed. In practice, alternative randomisation schemes are often used. The most common is stratified randomisation. It is well known that such schemes introduce a dependence in the data, which if ignored, leads to conservative inferences for treatment effect estimates \cite{Shao2010,Kahan2012,Bugni2016}. In particular, although stratified randomisation will not affect consistency of the estimators, it will affect their variance. In the following, we consider how stratified randomisation impacts on $\hat{\mu}_{1}(z)$, $\hat{\mu}_{2}(z)$,  and $\hat{\mu}_{3}(z)$. 

\subsection{$\hat{\mu}_{1}(z)$}
Stratified randomisation ensures that for the covariates used in the randomisation, their sample distribution is identical or near identical across treatment groups. This means that the distribution of these covariates in those randomised to $Z=z$ is, in expectation, closer to the population distribution than would be expected under simple randomisation. As such, we would expect the usual i.i.d. variance estimator for $\hat{\mu}_{1}(z)$ to be biased upwards when stratified randomisation is used. We investigate this conjecture in the simulation study described in Section \ref{simulations}.

\subsection{$\hat{\mu}_{2}(z)$}
We first consider $\hat{\mu}_{2}(z)$ when the covariates adjusted for include all of those used in the stratified randomisation. The theory developed previously for the variance of $\hat{\mu}_{2}(z)$ assumed that patients' data are i.i.d. In Appendix \ref{stratmuhat2} we show that provided the outcome model is correctly specified, the variance of $\hat{\mu}_{2}(z)$ can be estimated as described previously in the case of simple randomisation.

Now suppose that the covariates $X$ adjusted for are exactly those used in the stratified randomisation, i.e. that the outcome model does not adjust for any additional covariates not used in the stratified randomisation. As noted previously, under stratified randomisation, the sample distribution of $X$ is identical, or near identical, across the treatment groups. As such, we might expect that $\hat{\mu}_{2}(z)$ would not have improved efficiency compared to $\hat{\mu}_{1}(z)$, since those randomised to treatments other than $z$ give no additional information about the population distribution of $X$. To demonstrate this conjecture in a special case, suppose that there exists a single binary baseline covariate $X$, randomisation was performed stratified on $X$, and that there are two treatment groups. Suppose we fit a canonical GLM for the outcome conditional on $Y$ and $X$ and $Z$, with main effects of $X$ and $Z$. The outcome model estimating equations then consist of
\begin{eqnarray*}
\sum^{n}_{i=1} \{Y_{i} - h(X_{i},Z_{i},\hat{\beta}) \} \begin{pmatrix} 1 \\ X_{i} \\ Z_{i} \end{pmatrix} = 0
\end{eqnarray*}
A consequence is that $\sum Z_{i} Y_{i} = \sum Z_{i} h(X_{i},1,\hat{\beta})$. Since the sample distribution of $X$ is identical (or almost identical) across treatment groups as a consequence of stratified randomisation,
\begin{eqnarray*}
\frac{\sum^{n}_{i=1} Z_{i} h(X_{i},1,\hat{\beta})}{\sum^{n}_{i=1} Z_{i}} \approx \frac{\sum^{n}_{i=1} (1-Z_{i}) h(X_{i},1,\hat{\beta})}{\sum^{n}_{i=1} (1-  Z_{i})}
\end{eqnarray*}
so that
\begin{eqnarray*}
\hat{\mu}_{2}(1) &=& n^{-1} \sum^{n}_{i=1} h(X_{i},1,\hat{\beta}) \\
&\approx &  \frac{\sum^{n}_{i=1} Z_{i} h(X_{i},1,\hat{\beta})}{\sum^{n}_{i=1} Z_{i}} \\
&=&  \frac{\sum^{n}_{i=1} Z_{i} Y_{i}}{\sum^{n}_{i=1} Z_{i}} \\
&=& \hat{\mu}_{1}(1)
\end{eqnarray*}
A similar argument naturally carries through for $z=0$. Thus the estimators $\hat{\mu}_{1}(z)$ and $\hat{\mu}_{2}(z)$ are essentially equivalent, and there is no gain in precision over $\hat{\mu}_{1}(z)$.

\subsection{$\hat{\mu}_{3}(z)$}
In general, the variance estimator described assuming data are i.i.d. for $\hat{\mu}_{3}(z)$ is not expected to be valid under stratified randomisation. This can be seen by the fact that $\hat{\mu}_{3}(z)$ reduces to $\hat{\mu}_{1}(z)$ upon choosing $h(X,Z,\beta)=0$, and as described previously, we expect the i.i.d. variance estimator for $\hat{\mu}_{1}(z)$ to be biased upwards under stratified randomisation.

\section{Simulations}
\label{simulations}
In this section we present simulation results to investigate the finite sample properties of the methods described, first where random permuted block randomisation is used, and second where stratified randomisation is used. Treatment allocation was performed with $P(Z=1)=0.5$. We simulated data for trials of size $n=400$, with a single binary covariate $X \sim Bernoulli(0.5)$. For a random 25\% of patients their follow-up time $T$ was generated from a uniform distribution on $(0,1)$, with the remainder having follow-up $T=1$. We investigated the estimators described previously in four different scenarios:
\begin{enumerate}
	\item $Y_{i}$ generated from a Poisson distribution, with mean $ \gamma_{i} T_{i} \exp(3X_{i}+Z_{i})$, where $\gamma_{i}$ was gamma distributed with shape 2 and scale 1/2. The working model used for $\hat{\mu}_{2}(z)$ and $\hat{\mu}_{3}(z)$ was a negative binomial model, with main effects of $X_{i}$ and $Z_{i}$, and $\log(T_{i})$ as offset
	\item $Y_{i}$ generated in the same way, but with $\gamma_{i}$ a log normally distributed random effect with mean 1 and variance 1/2. The working model used for $\hat{\mu}_{2}(z)$ and $\hat{\mu}_{3}(z)$ was again a negative binomial model with main effects of $X_{i}$ and $Z_{i}$, and $\log(T_{i})$ as offset
	\item $Y_{i}$ generated from a Poisson distribution, with mean $ \gamma_{i} T_{i} \exp(3X_{i}+Z_{i}-1.5X_{i}Z_{i})$, where $\gamma_{i}$ was gamma distributed with shape 2 and scale 1/2. The working model used for $\hat{\mu}_{2}(z)$ and $\hat{\mu}_{3}(z)$ was again a negative binomial model with main effects of $X_{i}$ and $Z_{i}$, and $\log(T_{i})$ as offset
	\item $Y_{i}$ generated from a Poisson distribution, with mean $ \gamma_{i} T_{i} \exp(3X_{i}+Z_{i}-1.5X_{i}Z_{i})$, where $\gamma_{i}$ was gamma distributed with shape 2 and scale 1/2.  The working model used for $\hat{\mu}_{2}(z)$ and $\hat{\mu}_{3}(z)$ was a Poisson model with main effects of $X_{i}$ and $Z_{i}$, and $\log(T_{i})$ as offset 
\end{enumerate}
We note that these scenarios assume very strong associations between $X$ and the outcome, since it is the strength of this association that largely drives the differences we expect to see. The variance of $\hat{\mu}_{1}(z)$ was estimated using equation \eqref{ratemuhat1var}. The Qu and Luo variance estimator for $\hat{\mu}_{2}(z)$ was calculated using equation \eqref{QuLuoVar}, with a sandwich variance estimator used to estimate the variance of $\hat{\beta}$. The term in equation \eqref{quluo_adjust} was then added to this variance to allow for the fact that $X$ was not fixed in repeated samples. The variance of $\hat{\mu}_{3}(z)$ was calculated using equation \eqref{muhat3estvarfup}. Confidence intervals were calculated on the log scale and then back transformed. 

Table \ref{simresults1} shows the results of the simulations with random permuted block randomisation. As expected, $\hat{\mu}_{2}(1)$ was unbiased when the outcome regression model was correctly specified (scenario 1). It was also unbiased when the random effects distribution was modelled incorrectly (scenario 2), since as described earlier, with full data, the negative binomial model gives consistent estimates of the regression coefficients provided the conditional mean function is correctly specified. In scenario 3, the conditional mean function in the outcome model was misspecified, and so $\hat{\mu}_{2}(1)$ was biased. In scenario 4 $\hat{\mu}_{2}(1)$ was unbiased, despite the conditional mean function being misspecified, since a canonical GLM (Poisson) model was used. As expected, $\hat{\mu}_{2}(1)$ was more efficient than $\hat{\mu}_{1}(1)$. Confidence intervals based on the fixed $X$ standard error described by Qu and Luo \cite{Qu2015} had coverage below the nominal 95\% level, as predicted by theory. In contrast, in those scenarios where $\hat{\mu}_{2}(1)$ was unbiased, the coverage of the random $X$ confidence intervals was close to the 95\% level. This was the case even in scenario 4, using a misspecifed Poisson model, a result which does not appear to be implied theoretically. The robust estimator $\hat{\mu}_{3}(1)$ was unbiased in all four scenarios, was more efficient than $\hat{\mu}_{1}(1)$, and confidence intervals had coverage just sllightly below the 95\% level.

\begin{table}[ht]
\centering
\caption{Bias and coverage of 95\% confidence intervals for $\hat{\mu}_{1}(1)$, $\hat{\mu}_{2}(1)$, $\hat{\mu}_{3}(1)$, across 10,000 simulations. Rel. eff. denotes the ratio of the empirical variance of $\hat{\mu}_{1}(1)$ to the variance of $\hat{\mu}_{2}(1)$ or $\hat{\mu}_{3}(1)$.}
\begin{tabular}{lrrrr}
  \hline
 & Scenario 1 & Scenario 2 & Scenario 3 & Scenario 4 \\ 
  \hline
  $\hat{\mu}_{1}(1)$ \\
Mean  & 3.88 & 3.88 & 2.80 & 2.81 \\ 
  95\% CI Cov. & 94.53 & 94.28 & 94.69 & 94.61 \\ 
  \hline
    $\hat{\mu}_{2}(1)$ \\
  Bias  & 0.00 & 0.00 & 0.18 & 0.00 \\ 
  Rel. eff. & 1.28 & 1.28 & 1.14 & 1.22 \\ 
  Fixed $X$ 95\% CI Cov. & 89.61 & 89.20 & 81.96 & 91.15 \\ 
  Random $X$ 95\% CI Cov. & 94.41 & 94.20 & 88.87 & 95.08 \\ 
  \hline
    $\hat{\mu}_{3}(1)$ \\
  Bias  & 0.00 & 0.00 & 0.00 & 0.00 \\ 
  Rel. eff. & 1.26 & 1.25 & 1.21 & 1.22 \\ 
  95\% CI Cov. & 94.47 & 94.30 & 94.67 & 94.56 \\ 
   \hline
\end{tabular}
\label{simresults1}
\end{table}

Table \ref{simresults2} shows the results of simulations where randomisation was stratified on $X$. As conjectured, the i.i.d. confidence intervals for $\hat{\mu}_{1}(1)$ had coverage higher than 95\%. Also, since the covariate $X$ was stratified on in the randomisation, neither $\hat{\mu}_{2}(1)$ nor $\hat{\mu}_{3}(1)$ had improved efficiency relative to $\hat{\mu}_{1}(1)$. Fixed $X$ confidence intervals for $\hat{\mu}_{2}(1)$ again under covered, whereas the random $X$ intervals had correct coverage whenever $\hat{\mu}_{2}(1)$ was unbiased. Interestingly, confidence intervals for $\hat{\mu}_{3}(1)$ had correct coverage, even in scenario 3 where the working conditional mean model was misspecified. Further research is warranted to understand the reason for this.

\begin{table}[ht]
\centering
\caption{Bias and coverage of 95\% confidence intervals for $\hat{\mu}_{1}(1)$, $\hat{\mu}_{2}(1)$, $\hat{\mu}_{3}(1)$, across 10,000 simulations, with stratified randomisation. Rel. eff. denotes the ratio of the empirical variance of $\hat{\mu}_{1}(1)$ to the variance of $\hat{\mu}_{2}(1)$ or $\hat{\mu}_{3}(1)$.}
\begin{tabular}{lrrrr}
  \hline
 & Scenario 1 & Scenario 2 & Scenario 3 & Scenario 4 \\ 
  \hline
    $\hat{\mu}_{1}(1)$ \\
 Mean  & 3.88 & 3.88 & 2.81 & 2.81 \\ 
  95\% CI Cov. & 96.84 & 96.83 & 96.70 & 96.88 \\ 
  \hline
      $\hat{\mu}_{2}(1)$ \\
  Bias & 0.00 & 0.00 & 0.18 & 0.00 \\ 
  Rel. eff. & 1.02 & 1.02 & 0.94 & 1.00 \\ 
  Fixed $X$ 95\% CI Cov. & 89.52 & 89.44 & 82.28 & 91.88 \\ 
  Random $X$ 95\% CI Cov. & 94.19 & 94.39 & 88.65 & 95.21 \\ 
  \hline
      $\hat{\mu}_{3}(1)$ \\
  Bias $\hat{\mu}_{3}(1)$ & 0.00 & 0.00 & 0.00 & 0.00 \\ 
  Rel. eff. & 1.00 & 1.00 & 1.00 & 1.00 \\ 
  95\% CI Cov. & 94.05 & 94.27 & 94.78 & 94.90 \\ 
   \hline
\end{tabular}
\label{simresults2}
\end{table}

\section{Illustrative analysis}
\label{illustrative}
In this section we re-analyse data from the Sirocco trial, a randomised placebo controlled phase 3 trial designed to investigate the efficacy and safety of benralizumab in patients with severe asthma \cite{sirocco}. Across 374 sites in 17 countries, 1,205 patients were randomised. Randomisation was stratified by age group, country/region, and blood eosinophil count category. The primary analysis population consisted of those patients with blood eosinophil counts at least 300 cells per μL, in which 267 were randomised to placebo, 275 to benralizumab 30 mg every 4 weeks, and 267 to benralizumab 30 mg every 8 weeks. The analyses presented here are based on a de-identified version of the data with patients who revoked their informed consent removed, and which includes 257 patients on placebo, 263 patients on benra 4 week, and 256 patients on benra 8 week.

The primary outcome was asthma exacerbations during a planned 48 week follow-up period. In the subset used here, approximately 95\% of patients had at least 46 of the planned 48 weeks of follow-up. The primary outcome was analysed using negative binomial regression, with randomised treatment group, region, exacerbations in the previous year (two, three, or four or more), and oral corticosteroid use at time of randomisation. As part of the trial's results \cite{sirocco}, $\hat{\mu}_{2}(z)$ was calculated using the negative binomial model for each of the three treatment groups, with 95\% CIs calculated using the method proposed by \cite{Qu2015}.

Table \ref{siroccotab} shows the estimated asthma exacerbation rates using $\hat{\mu}_{1}(z)$, $\hat{\mu}_{2}(z)$ and $\hat{\mu}_{3}(z)$ for each of the three treatment groups. For $\hat{\mu}_{2}(z)$ and $\hat{\mu}_{3}(z)$ estimates are shown both for a negative binomial and Poisson model, in which the same covariates were adjusted for as in the trial's primary analysis. Confidence intervals are shown based on the method proposed by Qu and Luo \cite{Qu2015}, treating the baseline covariates as fixed, and using the adjustment given in equation \eqref{quluo_adjust}. Sandwich variances were used to estimate the variance covariance matrix of the models' regression parameters.

For the rate under placebo the covariate adjusted estimates are all smaller than the crude rate. For benra 4 weeks the crude and adjusted estimates are quite similar, while for benra 8 weeks the covariate adjusted estimates are all somewhat higher than the crude rate. Differences between the crude and adjusted estimates are to be expected since as described previously, the latter adjust the crude rates for chance imbalance in the baseline covariate distribution between randomised groups. Differences between the various covariate adjusted estimates for each treatment were small.

The confidence intervals for $\hat{\mu}_{2}(z)$ constructed as proposed by Qu and Luo \cite{Qu2015} were very slightly narrower than those that include the adjustment given in equation \eqref{quluo_adjust}. Thus here assuming the covariates were fixed for the purposes of variance estimation of $\hat{\mu}_{2}(z)$ made very little difference. This can be explained by the fact that although some of the baseline covariates adjusted for were statistically significant with moderately large associations, their estimated associations were nevertheless smaller in magnitude than what was assumed in the simulation study.

\begin{table}[ht]
\centering
\caption{Estimates of asthma exacerbation rate from the Sirocco trial with 95\% confidence intervals. NB - negative binomial. `Fixed X' and `Random X' correspond to confidence intervals calculated assuming the covariates are fixed or random respectively.}
\begin{tabular}{llll}
  \hline
 & Placebo & Benra 4 weeks & Benra 8 weeks \\ 
  \hline
$\hat{\mu}_{1}$ & 1.536 (1.287, 1.833) & 0.836 (0.667, 1.049) & 0.652 (0.523, 0.813) \\ 
  $\hat{\mu}_{2}$ NB Fixed X & 1.464 (1.245, 1.722) & 0.819 (0.561, 1.195) & 0.710 (0.456, 1.107) \\ 
  $\hat{\mu}_{2}$ NB Random X & 1.464 (1.240, 1.729) & 0.819 (0.560, 1.197) & 0.710 (0.455, 1.109) \\ 
  $\hat{\mu}_{3}$ NB & 1.483 (1.247, 1.764) & 0.839 (0.662, 1.063) & 0.676 (0.546, 0.836) \\ 
  $\hat{\mu}_{2}$ Poisson Fixed X & 1.490 (1.268, 1.751) & 0.841 (0.575, 1.228) & 0.676 (0.413, 1.107) \\ 
  $\hat{\mu}_{2}$ Poisson Random X & 1.490 (1.263, 1.758) & 0.841 (0.574, 1.230) & 0.676 (0.412, 1.108) \\ 
  $\hat{\mu}_{3}$ Poisson & 1.482 (1.247, 1.762) & 0.840 (0.662, 1.065) & 0.674 (0.545, 0.834) \\ 
   \hline
\end{tabular}
\label{siroccotab}
\end{table}

\section{Discussion}
\label{discussion}
When a randomised trial's analysis adjusts for baseline covariates, we believe it is advisable to also estimate and report baseline adjusted estimates of mean outcome under each treatment. Estimating the mean outcome when the baseline covariates are equal to their corresponding means arguably makes little sense when some baseline covariates are categorical. Instead, the marginal mean estimator $\hat{\mu}_{2}(z)$ proposed by Qu and Luo \cite{Qu2015}, targets the same parameter as the unadjusted group mean $\hat{\mu}_{1}(z)$, but removes variation attributable to chance imbalance in the distribution of baseline covariates between randomised groups. Assuming the outcome model used is correctly specified, this leads to mean estimates with improved precision.

We have shown that the variance estimator for $\hat{\mu}_{2}(z)$ described by Qu and Luo \cite{Qu2015} will in general be biased downwards when, as is the case usually, patients' baseline covariate values would not be fixed in repeated sampling. This bias may however be small unless the covariate effects on outcome are large. We have proposed a simple adjustment to the Qu and Luo variance estimator, which results in a valid variance estimator provided the outcome model is correctly specified.

Semiparametric methods which adjust for baseline covariates offer the opportunity for tests of treatment effect with improved power while remaining valid when the working model involved is incorrectly specified \cite{Zhang2008}. As we have shown, this methodology leads directly also to estimates of $\mu(z)$, offering improved precision relative to $\hat{\mu}_{1}(z)$, but unlike $\hat{\mu}_{2}(z)$, retaining consistency even when the working model is misspecified.

As others have described previously, use of covariate adaptive randomisation schemes such as stratified randomisation offers the potential for more precise inferences to be made. Nevertheless, this potential is only realised when the analysis correctly accounts for the randomisation scheme. As we have described, stratified randomisation causes naive i.i.d. variance estimates for $\hat{\mu}_{1}(z)$ to be upwardly biased. Further research is warranted to understand under what conditions the i.i.d. variance estimator for $\hat{\mu}_{3}(z)$ remains valid under covariate adaptive randomisation schemes. 

We have throughout assumed that the baseline covariates and outcome are fully observed. While the former are typically fully observed in randomised trials, the latter is often incomplete. When some outcomes are missing in those randomised to $Z=z$, $\hat{\mu}_{1}(z)$ is only unbiased if missingness is independent of outcome within each treatment group, an assumption which often is not plausible. When the outcome model used is correctly specified, the adjusted estimator $\hat{\mu}_{2}(z)$ is however valid under the weaker assumption that missingness is independent of outcome conditional on covariates. For $\hat{\mu}_{3}(z)$, unbiasedness is only guaranteed in general when missingness is independent of outcome. We note that the methods described here could of course be used in conjunction with multiple imputation or inverse probability weighting for handling such missingness.

\section{Acknowledgements}
The author thanks Stijn Vansteelandt for helpful comments and discussion on the asymptotic theory. He also thanks the Sirocco trial study team for facilitating use of the data for the illustrative analysis.

\appendix

\section{Asymptotic theory for $\hat{\mu}_{2}(z)$}
\subsection{Consistency}
\label{muhat2bias}
Here we show that $\hat{\mu}_{2}(z)$ is consistent for certain outcome models, irrespective of whether the outcome model is correctly specified. As noted in Section \ref{baselineadj}, $\hat{\mu}_{2}(z)$ is consistent if and only if $E\{h(X,z,\tilde{\beta})\}=\mu(z)$. Suppose that the outcome model is a canonical generalised linear model. Then if there are $k$ treatment groups, the estimating equations for $\beta$ are
\begin{eqnarray*}
\sum^{n}_{i=1} \{Y_{i} - h(X_{i},Z_{i},\beta) \} \begin{pmatrix} 1 \\ X_{i} \\ 1(Z_{i}=1) \\ \vdots \\ 1(Z_{i}=k) \end{pmatrix} = 0
\end{eqnarray*}
and the large sample limits $\tilde{\beta}$ satisfies
\begin{eqnarray*}
E \left[ \{Y - h(X,Z,\tilde{\beta}) \} \begin{pmatrix} 1 \\ X \\ 1(Z=1) \\ \vdots \\ 1(Z=k) \end{pmatrix} \right] = 0
\end{eqnarray*}
This means in particular that 
\begin{eqnarray*}
0 &=& E \left[ 1(Z=z) \{Y - h(X,Z,\tilde{\beta}) \}  \right] \\
&=& \pi_{z}\left[\mu(z) - E\{h(X,z,\tilde{\beta})\} \right]
\end{eqnarray*}
so that $E\{h(X,z,\tilde{\beta})\}=\mu(z)$, and hence $\hat{\mu}_{2}(z)$ is consistent.

Now consider the setting where $Y$ denotes the number of events occurring for a patient, and $T$ denotes their follow-up time. Suppose we use Poisson regression with the canonical log link, with $\log(T)$ as an offset. Then we have that
\begin{eqnarray*}
0 &=& E \left[ 1(Z=z) \{Y - T h(X,Z,\tilde{\beta}) \}  \right]
\end{eqnarray*}
Now suppose that $T \ci X | Z$. Then
\begin{eqnarray*}
0 &=& \pi_{z}E(T|Z=z)\mu(z) - \pi_{z} E(T|Z=z)E\{h(X,z,\tilde{\beta})\}
\end{eqnarray*}
so that $E\{h(X,z,\tilde{\beta})\}=\mu(z)$, and hence $\hat{\mu}_{2}(z)$ is again consistent.

\subsection{Variance}
\label{appVariance}
In this appendix we derive an expression for the sampling variance of $\hat{\mu}_{2}(z)$, assuming the baseline covariates are random in repeated sampling. To find the asymptotic distribution of $\hat{\mu}_{2}(z)$ which accounts for $X$ being random, we use the estimating equation theory described by Newey and McFadden \cite{Newey1994}. In addition to the parameters $\beta$ indexing the conditional mean function, the outcome model may contain additional parameters $\eta$. We let $\theta=(\beta,\eta)$ denote the combined parameter. We assume that $\theta$ is estimated by the value $\hat{\theta}=(\hat{\beta},\hat{\eta})$ solving a set of estimating equations
\begin{eqnarray}
0=\sum^{n}_{i=1} m(Y_{i},X_{i},Z_{i},\hat{\theta})
\end{eqnarray}
for some estimating function $m(Y,X,Z,\theta)$. In the case where the outcome model is fitted by maximum likelihood, the estimating equations correspond to the likelihood score equations. We let $\tilde{\theta}=(\tilde{\beta},\tilde{\eta})$ denote the large sample probability limit of the estimator.

The estimator $\hat{\mu}_{2}(z)$ is a so called two step estimator. Suppose that $E\{h(X,z,\tilde{\beta})\}=\mu(z)$, so that $\hat{\mu}_{2}(z)$ is consistent. Application of Theorem 6.1 of Newey and McFadden \cite{Newey1994} then states that, under regularity conditions, $\hat{\mu}_{2}(z)$ is asymptotically normally distributed, with mean $\mu(z)$, and variance
\begin{eqnarray}
n^{-1} \Var \left[  h(X,z,\tilde{\beta}) - \mu(z) +G_{\theta} \psi(Y,X,Z)\right] \label{nonlinAsyVar}
\end{eqnarray}
where $\psi(Y,X,Z)$ is the so called influence function of $\hat{\theta}$, and 
\begin{eqnarray}
G_{\theta} &=& E \left[\frac{\partial}{\partial \theta^{T}} \{h(X,z,\tilde{\beta}) - \mu(z)  \} \right] \nonumber \\
&=& \begin{pmatrix} E \left[  \frac{\partial}{\partial \beta^{T}} h(X,z,\tilde{\beta})  \right] & 0 \end{pmatrix} \nonumber \\
&=& \begin{pmatrix} G_{\beta} & 0 \end{pmatrix} \label{Gbeta}
\end{eqnarray}
The influence function of $\hat{\theta}$ is equal to
\begin{eqnarray*}
\psi(Y_{i},X_{i},Z_{i}) &=& - M^{-1}  m(Y_{i},X_{i},Z_{i},\tilde{\theta})
\end{eqnarray*}
where
\begin{eqnarray*}
M &=& E \left[ \frac{\partial}{\partial \theta^{T}}  m(Y,X,Z,\tilde{\theta})  \right] \
\end{eqnarray*}
In full generality, the variance of $\hat{\mu}_{2}(z)$ can be estimated by
\begin{eqnarray}
\widehat{\Var}\{\hat{\mu}_{2}(z)\} = 
n^{-2} \sum^{n}_{i=1} \left[ h(X,z,\hat{\beta}) - \hat{\mu}(z) +\hat{G}_{\theta} \hat{\psi}(Y,X,Z)  \right]^{2} \label{muhat2varest}
\end{eqnarray}
where 
\begin{eqnarray*}
\hat{\psi}(Y,X,Z) = - \hat{M}^{-1}  m(Y_{i},X_{i},Z_{i},\hat{\theta}),
\end{eqnarray*}
\begin{eqnarray*}
\hat{M} = n^{-1} \sum^{n}_{i=1}  \frac{\partial}{\partial \theta^{T}} m(Y_{i},X_{i},Z_{i},\hat{\theta}),
\end{eqnarray*}
and
\begin{eqnarray*}
\hat{G}_{\theta} = \begin{pmatrix} n^{-1}  \sum^{n}_{i=1}\frac{\partial h(X_{i},z,\hat{\theta})}{\partial \theta^{T}} & 0 \end{pmatrix}
\end{eqnarray*}

In certain situations, a simplification of the preceding variance estimator is possible. The variance in equation \eqref{nonlinAsyVar} can be expanded as
\begin{multline}
\Var \left[  h(X,z,\tilde{\beta}) - \mu(z) +G_{\theta} \psi(Y,X,Z)\right]   \\
= \Var \left[h(X,z,\tilde{\beta}) - \mu(z) \right] + \Var \left[ G_{\theta} \psi(Y,X,Z) \right] \\
+ 2 \Cov \left[ h(X,z,\tilde{\beta}) - \mu(z), G_{\theta} \psi(Y,X,Z) \right] \label{varexpansion}
\end{multline}
The covariance term can be expanded as
\begin{eqnarray*}
\Cov \left[ E \{  h(X,z,\tilde{\beta}) - \mu(z) | X,Z \}, E \{G_{\theta} \psi(Y,X,Z) | X,Z\}  \right] \\
+ E \left[\Cov\{h(X,z,\tilde{\beta}) - \mu(z), G_{\theta} \psi(Y,X,Z)  |X,Z\} \right]
\end{eqnarray*}
The second term in this expression is zero, since the first component is constant, conditional on $X$ and $Z$. Now suppose that the outcome model estimating function is conditionally unbiased, in the sense that
\begin{eqnarray}
E\{m(Y,X,Z,\tilde{\theta})|X,Z\}=0
\end{eqnarray}
This condition typically holds if the outcome model is a correctly specified parametric regression model estimated by maximum likelihood, where the estimating equation for $\theta$ is the likelihood score equation (see for example page 393 of \cite{Wooldridge2002}). It holds for generalised linear models provided the conditional mean function is correctly specified. It is also satisfied when the outcome model is a correctly specified semiparametric conditional mean model \cite{Rotnitzky/Robins:1997}. When the condition holds, it then follows that $E \{G_{\theta} \psi(Y,X,Z) | X,Z\}=0$, in which case the covariance term in \eqref{varexpansion} is zero. 

Now consider the case where the outcome model is negative binomial regression, and suppose the conditional mean function is correctly specified. In this case the part of the estimating equations corresponding to $\beta$ have expectation conditional on $X$ and $Z$ equal to zero. It also follows (see \cite{Cameron2013}) that the matrix $M$ is block diagonal, and hence so is $M^{-1}$. Then we have
\begin{eqnarray*}
E \{G_{\theta} \psi(Y,X,Z) | X,Z\} &=&  - \begin{pmatrix} G_{\beta} & 0 \end{pmatrix} M^{-1}  E\left\{m(Y,X,Z,\tilde{\theta}|X,Z) \right\} \\
&=& - \begin{pmatrix} G_{\beta} & 0 \end{pmatrix} \begin{pmatrix}m^{-1}_{11}  & 0 \\ 0 & m^{-1}_{22} \end{pmatrix}  \begin{pmatrix} 0 \\ \phi(X,Z) \end{pmatrix} \\
&=& 0
\end{eqnarray*}
Thus in this case again the covariance term in \eqref{varexpansion} is zero. 

Suppose then that the covariance in equation \eqref{varexpansion} is zero. A key property of the influence function $\psi(Y,X,Z)$ is that $\hat{\theta}=(\hat{\beta},\hat{\eta})$ is asymptotically normally distributed with mean $\tilde{\theta}=(\tilde{\beta},\tilde{\eta})$ and variance $n^{-1}\Var\{\psi(Y,X,Z)\}$. Consequently,
\begin{eqnarray*}
\Var \{ G_{\theta} \psi(Y,X,Z) \} 
&=& n G_{\theta} \Var(\hat{\theta}) G^{T}_{\theta} \\
&=& n \begin{pmatrix} G_{\beta} & 0 \end{pmatrix} \Var(\hat{\theta})  \begin{pmatrix} G^{T}_{\beta}\\ 0 \end{pmatrix} \\
&=& n G_{\beta} \Var(\hat{\beta})  G^{T}_{\beta}
\end{eqnarray*}
We then have that $\hat{\mu}_{2}(z)$ has variance
\begin{eqnarray}
n^{-1}  \Var \left\{ h(X,z,\tilde{\beta}) - \mu(z) \right\} + G_{\beta}  \Var(\hat{\beta}) G^{T}_{\beta} 
\label{varcorrectOM}
\end{eqnarray}
The second term in this expression corresponds to the variance estimator proposed by Qu and Luo \cite{Qu2015}. As such, this expression shows that when the covariance term in \eqref{varexpansion} is zero, the variance estimator proposed by Qu and Luo underestimates the variance of $\hat{\mu}_{2}(z)$ by $n^{-1} \Var \left[ h(X,z,\tilde{\beta})- \mu(z) \right]$. Although this bias term tends to zero as $n$ increases, it converges at the same rate as the other contribution to the variance, such that the Qu and Luo variance estimator will be biased downwards even for large sample sizes. Equation \eqref{varcorrectOM} implies that a valid variance estimate for $\hat{\mu}_{2}(z)$ can be obtained, assuming the outcome model estimating function is conditionally unbiased, by adding 
\begin{eqnarray*}
n^{-2} \sum^{n}_{i=1}  \{ h(X,z,\hat{\beta}) - \hat{\mu}(z) \}^{2}
\end{eqnarray*}
to the Qu and Luo variance estimator given in equation \eqref{QuLuoVar}.

\subsection{Efficiency compared to $\hat{\mu}_{1}(z)$}
\label{muhat2eff}
In this section we give a sketch proof to show that $\hat{\mu}_{2}(z)$ is a more efficient estimator than $\hat{\mu}_{1}(z)$, when the outcome model is a correctly specified parametric model estimated by maximum likelihood. Let $f(Y|X,Z,\theta)$ denote this model, with $E(Y|X,Z)=h(X,Z,\theta)$. We assume a nonparametric model for $X$, and $P(Z=z|X)=\pi_{z}$ by randomisation. We then use semiparametric theory as described in \cite{Tsiatis:2006} to demonstrate that $\hat{\mu}_{2}(z)$ is the most efficient estimator in this semiparametric model, and therefore that it is more efficient than $\hat{\mu}_{1}(z)$.

To derive the efficient estimator, we find the tangent space. Thus consider a parametric submodel, where $f(X|\lambda)$ is a parametric model for $X$. Since no model is required for $P(Z|X)$, the tangent space for this (arbitrary) parametric submodel is $\mathcal J_{\theta} \oplus \mathcal J_{\lambda}$ where 
\begin{eqnarray*}
\mathcal J_{\theta} = \left \{ B^{1 \times p} S_{\theta}(Y,X,Z) : \mbox{for all vectors } B^{1 \times p} \right \}
\end{eqnarray*}
where $S_{\theta}(Y,X,Z)$ is the score vector corresponding to the parametric model $f(Y|X,Z,\theta)$, and
\begin{eqnarray*}
\mathcal J_{\lambda} = \left \{ B^{1 \times q} S_{\lambda}(X) : \mbox{for all vectors } B^{1 \times q} \right \}
\end{eqnarray*}
where $S_{\lambda}(X)$ is the score vector corresponding to the (arbitrary) parametric model for $X$. By definition, the tangent space for the semiparametric model is the mean square closure of all parametric submodel tangent spaces. Since $\theta$ and $\lambda$ are variationally independent, the tangent space for the semiparametric model is $\mathcal J_{\theta} \oplus \mathcal J_{1}$ where 
\begin{eqnarray*}
\mathcal J_{1} =  \left \{ g(X): E(g(X))=0 \right\}
\end{eqnarray*}
is the mean square closure of the submodel tangent spaces $\mathcal J_{\lambda}$. The estimator $\hat{\mu}_{1}(z)$ clearly belongs to the class of RAL estimators for this semiparametric model. The influence function for $\hat{\mu}_{1}(z)$ is
\begin{eqnarray*}
\frac{1(Z=z)}{\pi_{z}} \left\{ Y - \mu(z) \right\}
\end{eqnarray*}
The influence function of $\hat{\mu}_{2}(z)$ is
\begin{eqnarray*}
h(X,z,\theta) - \mu(z) + G_{\theta} \{ E(S_{\theta} S^{T}_{\theta}) \}^{-1} S_{\theta}(Y,X,Z)
\end{eqnarray*}
where 
\begin{eqnarray*}
G_{\theta} = E \left[\frac{\partial h(X,z,\theta)}{\partial \theta^{T}} \right] = E \left[\frac{\partial E(Y|X,z,\theta)}{\partial \theta^{T}} \right]
\end{eqnarray*}

We now show that $\hat{\mu}_{2}(z)$ is semiparametric efficient in this model. To show this, we show that the projection of the influence function of $\hat{\mu}_{1}(z)$ onto the tangent space is equal to the influence function of $\hat{\mu}_{2}$. Since the tangent space is the direct sum of two orthogonal spaces, the projection is equal to the sum of the projection onto $\mathcal J_{\theta}$ and the projection onto $\mathcal J_{1}$. The projection onto $\mathcal J_{1}$ is equal to
\begin{eqnarray*}
&E \left[\frac{1(Z=z)}{\pi_{z}} \left\{ Y - \mu(z) \right\} | X \right] = h(X,z,\theta) - \mu(z)
\end{eqnarray*}
The projection onto $\mathcal J_{\theta}$ is equal to
\begin{eqnarray*}
E\left[\frac{1(Z=z)}{\pi_{z}} \left\{ Y - \mu(z) \right\} S^{T}_{\theta} \right] \{ E(S_{\theta} S^{T}_{\theta}) \}^{-1} S_{\theta}(Y,X,Z)
\end{eqnarray*}
Assuming that $E(S_{\theta}|X,Z)=0$, the first expectation in this projection can be expanded as
\begin{eqnarray*}
E\left[\frac{1(Z=z)}{\pi_{z}} \left\{ Y - \mu(z) \right\} S^{T}_{\theta} \right] &=& E(Y S^{T}_{\theta}|Z=z) \\
&=& E\left[ E(Y S^{T}_{\theta}|X,Z=z)|Z=z \right] \\
&=& E \left[ \int Y S^{T}_{\theta} f(Y|X,z) dY | Z=z \right] \\
&=& E \left[ \int Y \frac{\frac{\partial}{\partial \theta^{T}} f(Y|X,z,\theta)}{f(Y|X,z,\theta)} f(Y|X,z,\theta) dY |Z=z \right] \\
&=& E \left[ \int Y \frac{\partial}{\partial \theta^{T}} f(Y|X,z,\theta) dY |Z=z \right] \\
&=& E \left[\frac{\partial}{\partial \theta^{T}} E(Y|X,z,\theta)  \right] \\
&=& G_{\theta}
\end{eqnarray*}
assuming sufficient regularity conditions in order to exchange differentiation and integration. We thus conclude that $\hat{\mu}_{2}(z)$ is semiparametric efficient. Lastly, since $\hat{\mu}_{2}(z)$ is semiparametric efficient for this model, it is in particular more efficient than $\hat{\mu}_{1}(z)$.

\section{Asymptotic theory for $\hat{\mu}_{3}(z)$}
\subsection{Derivation of $\hat{\mu}_{3}(z)$}
\label{appmuhat3}
To apply the results of Zhang \etal \cite{Zhang2008}, we first recall that the estimator $\hat{\mu}_{1}(z)$ which does not exploit $X$ can be expressed as solving estimating equations with estimating function $1(Z=z)\{Y-\mu(z)\}$. For this estimating function, application of their results then gives that all unbiased estimating functions for $\mu(z)$ may be written as
\begin{eqnarray*}
1(Z=z)\{Y-\mu(z)\} - \sum^{k}_{g=1} \{1(Z=g)-\pi_{g}\} a_{g}(X)
\end{eqnarray*}
where $k$ is the number of treatment groups and $a_{g}(X)$ is an arbitrary one dimensional function of $X$. The choice of $a_{g}(X)$ affects the efficiency of the resulting estimator of $\mu(z)$. The results of Zhang \etal show that the most efficient choice is
\begin{eqnarray*}
a_{g}(X)=E\{1(Z=z)\{Y-\mu(z)\}  | X,Z=g)
\end{eqnarray*}
which equals $0$ for $g \neq z$ and equals $E(Y|X,Z=z)-\mu(z)$ for $g=z$. Thus the efficient estimator has estimating function
\begin{eqnarray}
1(Z=z)\{Y-\mu(z)\} - \{1(Z=z)-\pi_{z}\}\{E(Y|X,Z=z)-\mu(z)\} 
\label{effestfun3}
\end{eqnarray}
Since $E(Y|X,Z)$ is unknown, we may postulate and fit a parametric working model for it, $h(X,Z,\beta)$, and replace $E(Y|X,Z)$ by $h(X,Z,\hat{\beta})$ in \eqref{effestfun3}. The resulting estimator of $\mu(z)$ is locally efficient, in the sense that if the working model is correctly specified, the estimator attains the semiparametric efficiency bound. Even if it is not correctly specified, the estimator remains consistent.

In a randomised trial, the randomisation probabilities $\pi_{z}$ are known by design. Nonetheless, it turns out (P206 of \cite{Tsiatis:2006}) that in general a more efficient estimator can be obtained by replacing the known $\pi_{z}$ by their empirical estimates $\hat{\pi}_{z}$, giving an estimating function
\begin{eqnarray*}
1(Z=z)\{Y-\mu(z)\} - \{1(Z=z)-\hat{\pi}_{z}\}\{E(Y|X,Z=z)-\mu(z)\}
\end{eqnarray*}
Solving the resulting estimating equation leads to
\begin{eqnarray*}
\hat{\mu}_{3}(z) &=& \hat{\mu}_{1}(z) - n^{-1} \sum^{n}_{i=1} \left[ \frac{1(Z_{i}=z)-\hat{\pi}_{z}}{\hat{\pi}_{z}} h(X_{i},z,\hat{\beta}) \right]
\end{eqnarray*}
The estimator $\hat{\pi}_{z}$ has estimating function $1(Z=z)-\pi_{z}$. As a so called two step estimator, it follows from Theorem 6.1 of Newey \cite{Newey1994} that $\hat{\mu}_{3}$ has asymptotic variance
\begin{eqnarray*}
\pi^{-2}_{z} E\left[\left\{1(Z=z)\{Y-\mu(z)\} - \{1(Z=z)-\pi_{z}\}\{h(X,z,\tilde{\beta})-\mu(z)\} +  \{1(Z=z)-\pi_{z}\}\{E(h(X,z,\tilde{\beta}))-\mu(z)\} \right\}^{2} \right] \\
= \pi^{-2}_{z} E\left[\left\{1(Z=z)\{Y-\mu(z)\} - \{1(Z=z)-\pi_{z}\}\{h(X,z,\tilde{\beta})-E(h(X,z,\tilde{\beta}))\} \right\}^{2} \right]
\end{eqnarray*}
It follows that the variance of $\hat{\mu}_{3}(z)$ can be estimated by
\begin{eqnarray*}
\hat{\pi}^{-2}_{z} n^{-2} \sum^{n}_{i=1} \left[1(Z_{i}=z)\{Y_{i}-\hat{\mu}_{3}(z)\} - \{1(Z_{i}=z)-\hat{\pi}_{z}\}\{h(X_{i},z,\hat{\beta})-\hat{\mu}_{2}(z)\} \right]^{2} 
\end{eqnarray*}

Now consider the setting where $Y$ denotes the number of events occurring for a patient, and $T$ denotes their follow-up time. The estimating function corresponding to $\hat{\mu}_{1}(z)$ is then $1(Z=z)\{Y-\mu(z)T\}$. Here $T$ and $Y$ are jointly considered as the outcome. Applying the results of Zhang \etal \cite{Zhang2008}, the efficient estimating function is
\begin{eqnarray*}
1(Z=z)\{Y-\mu(z)T\} - \{1(Z=z)-\pi_{z}\}\{E(Y|X,Z=z)-\mu(z)E(T|X,Z=z)\} 
\end{eqnarray*}
The two conditional expectations involved are again unknown, but we can construct a feasible estimator by postulating working models for them. To motivate our choices, we make the working assumption that $T \ci X | Z$, since if this did not hold, in general it would be the case that $E(Y|T,Z) \neq \mu(z)T$, such that $\hat{\mu}_{1}(z)$ would not be consistent. Under this conditional independence assumption, $E(T|X,Z=z)=E(T|Z=z)$, and this latter quantity can be estimated by the sample mean of $T$ in those randomised to $Z=z$, which we denote $\hat{\tau}_{z}$. For $E(Y|X,Z=z)$ we approximate it by $\hat{\tau}_{z} h(X,z,\beta)$, where $h(X,Z,\beta)$ denotes a working model prediction for $E(Y|X,Z,T=1)$. Again substituting $\hat{\pi}_{z}$ in place of $\pi_{z}$, this leads to
\begin{eqnarray*}
\hat{\mu}_{3}(z) &=& \frac{\sum^{n}_{i=1} 1(Z_{i}=z)Y_{i} - \{1(Z_{i}=z)-\hat{\pi}_{z}\}\{h(X_{i},z,\hat{\beta}) \hat{\tau}_{z}\}}{\sum^{n}_{i=1} 1(Z_{i}=z)T_{i}} \\
&=& \hat{\mu}_{1}(z) - n^{-1} \sum^{n}_{i=1} \left[ \frac{1(Z_{i}=z)-\hat{\pi}_{z}}{\hat{\pi}_{z}} h(X_{i},z,\hat{\beta}) \right]
\end{eqnarray*}
Again using Theorem 6.1 of Newey \cite{Newey1994}, it follows after further algebra that the variance of $\hat{\mu}_{3}(z)$ can be estimated by
\begin{eqnarray*}
\hat{\pi}^{-2}_{z} \hat{\tau}^{-2}_{z} n^{-2} \sum^{n}_{i=1} \left[1(Z_{i}=z)\{Y_{i}-\hat{\mu}_{3}(z)\} - \hat{\tau}_{z} \{1(Z_{i}=z)-\hat{\pi}_{z}\}\{h(X_{i},z,\hat{\beta})-\hat{\mu}_{2}(z)\} \right]^{2} 
\end{eqnarray*}

\subsection{Canonical generalised linear models}
\label{canonicalGLM}
We now show that for certain outcome model specifications, the estimator $\hat{\mu}_{2}(z)$ described in Section \ref{baselineadj} actually corresponds to an estimator of the form of $\hat{\mu}_{3}(z)$, and as such, inherits the asymptotic properties of the latter. First, we re-express $\hat{\mu}_{3}(z)$ as
\begin{eqnarray*}
\hat{\mu}_{3}(z) &=&  n^{-1} \sum^{n}_{i=1} \left[ h(X_{i},z,\hat{\beta}) + \frac{1(Z_{i}=z)}{\hat{\pi}_{z}} \left\{Y_{i} - h(X_{i},z,\hat{\beta}) \right\} \right]
\end{eqnarray*}
Now suppose that the estimating equations used to estimate $\beta$ include
\begin{eqnarray}
\sum^{n}_{i=1} 1(Z_{i}=z) \left\{ Y_{i} - h(X_{i},z,\hat{\beta}) \right\} = 0
\label{sampleMomentCond}
\end{eqnarray}
Then we have that $\hat{\mu}_{3}(z)=n^{-1} \sum^{n}_{i=1} h(X_{i},z,\hat{\beta})=\hat{\mu}_{2}(z)$. In this case, $\hat{\mu}_{2}(z)$ is consistent for $\mu(z)$, without requiring the outcome model to be correctly specified in any respect. Moreover, the variance estimator given in equation \eqref{muhat3estvar} can be used to estimate its variance, and this is valid even when the outcome model used is misspecified.

An important collection of models satisfying the condition given in equation \eqref{sampleMomentCond} are canonical link generalised linear models (GLMs) which include an intercept and main effect of $Z$ \cite{Cameron2013}. These include logistic regression for binary outcomes and Poisson regression for count outcomes.

\section{Stratified randomisation}
\label{stratmuhat2}
In this appendix we consider the impact of stratified randomisation on the variance of $\hat{\mu}_{2}(z)$, when the covariates used in the stratified randomisation are included within the covariates $X$. Then a reasonable assumption is that for $i\neq j$, $Y_{i} \ci (Y_{j},X_{j},Z_{j}) | (X_{i},Z_{i})$. That is, that conditional on a patient's covariates and treatment assignment, their outcome does not depend on the data from any other patients. The covariance between the contributions to the estimating equation for $\beta$ from any two patients is then
\begin{multline*}
\Cov\left[m(Y_{i},X_{i},Z_{i},\beta), m(Y_{j},X_{j},Z_{j},\beta)\right]  \\
 = \Cov\left[ E\{m(Y_{i},X_{i},Z_{i},\beta)|X_{i},Z_{i}\}, E\{m(Y_{j},X_{j},Z_{j},\beta)|X_{i},Z_{i}\} \right] \\
 +E\left[ \Cov\{m(Y_{i},X_{i},Z_{i},\beta), m(Y_{j},X_{j},Z_{j},\beta)|X_{i},Z_{i}\} \right]
\end{multline*}
If the outcome model is correctly specified, $E\{m(Y_{i},X_{i},Z_{i},\beta)|X_{i},Z_{i}\}=0$, and so the first covariance term will be zero. The conditional covariance in the second term is zero due to the stated conditional independence assumption. We have thus shown that under the stated conditions, the contributions to the estimating function for $\beta$ from distinct patients remain uncorrelated under stratified randomisation. Conditional on $\beta$, $\hat{\mu}_{2}(z)$ only depends on the data through $X$, and since $X$ remain i.i.d. under stratified randomisation, the variance estimator derived previously for $\hat{\mu}_{2}(z)$ given in equation \eqref{varcorrectOM} remains valid under stratified randomisation, under the stated conditions.

\end{document}